\documentclass[aps,prl,superscriptaddress,showpacs,floatfix,amsmath,amssymb,twocolumn]{revtex4}
\usepackage{epsfig}
\usepackage{bm}% bold math
\usepackage{hyperref}
\usepackage{balance}
\usepackage[usenames]{color}
\newcommand{\rms}{r_{nn}^{\rm rms}}
\def\USC{IGFAE, Instituto Galego de F\'{i}sica de Altas Enerx\'{i}as, Universidade de Santiago de Compostela, 15782 Santiago de Compostela, Spain}
\begin{document}

\title{%Decay of neutron pairs suddenly promoted to the continuum? \\
       %Probing strong neutron-neutron correlations near the neutron dripline? \\
       %Strong neutron-neutron correlations outside a nuclear core? \\
       {\boldmath{}Strong neutron pairing in core+$4n$ nuclei}}

\author{A.~Revel} 
\affiliation{Grand Acc\'el\'erateur National d'Ions Lourds (GANIL),
CEA/DRF-CNRS/IN2P3, Bvd Henri Becquerel, 14076 Caen, France}
\author{F.M.~Marqu\'es}
\affiliation{LPC Caen,
ENSICAEN, Universit\'e de Caen, CNRS/IN2P3, F-14050 CAEN Cedex, France}
\author{O.~Sorlin} 
\affiliation{Grand Acc\'el\'erateur National d'Ions Lourds (GANIL),
CEA/DRF-CNRS/IN2P3, Bvd Henri Becquerel, 14076 Caen, France}

\author{T.~Aumann}
\affiliation{Institut f\"ur Kernphysik, Technische Universit\"at Darmstadt, 64289 Darmstadt, Germany}
\affiliation{GSI Helmholtzzentrum f\"ur Schwerionenforschung, 64291 Darmstadt, Germany}
\author{C.~Caesar}
\affiliation{Institut f\"ur Kernphysik, Technische Universit\"at Darmstadt, 64289 Darmstadt, Germany}
\affiliation{GSI Helmholtzzentrum f\"ur Schwerionenforschung, 64291 Darmstadt, Germany}
\author{M.~Holl}
\affiliation{Institut f\"ur Kernphysik, Technische Universit\"at Darmstadt, 64289 Darmstadt, Germany} 
\author{V.~Panin}
\affiliation{Institut f\"ur Kernphysik, Technische Universit\"at Darmstadt, 64289 Darmstadt, Germany} 
\author{M.~Vandebrouck} 
\affiliation{Irfu, CEA, Universit\'e Paris-Saclay, 91191 Gif-sur-Yvette, France}
\author{F.~Wamers}
\affiliation{Institut f\"ur Kernphysik, Technische Universit\"at Darmstadt, 64289 Darmstadt, Germany}
\affiliation{GSI Helmholtzzentrum f\"ur Schwerionenforschung, 64291 Darmstadt, Germany}

\author{H.~Alvarez-Pol}\affiliation{\USC}
% ...

\collaboration{R$^3$B collaboration}

\begin{abstract}
The emission of neutron pairs from the neutron-rich $N\!=\!12$ isotones $^{18}$C and $^{20}$O has been studied by high-energy nucleon knockout from $^{19}$N and $^{21}$O secondary beams, populating unbound states of the two isotones up to 15~MeV above their two-neutron emission thresholds. 
The analysis of triple fragment-$n$-$n$ correlations shows that the decay
$^{19}$N$(-1p)^{18}$C$^*\!\rightarrow^{16}$C+$n$+$n$ is clearly dominated by direct pair emission. The two-neutron correlation strength, the largest ever observed, suggests the predominance of a $^{14}$C core surrounded by four valence neutrons arranged in strongly correlated pairs.
On the other hand, a significant competition of a sequential branch is found in the decay $^{21}$O$(-1n)^{20}$O$^*\!\rightarrow^{18}$O+$n$+$n$, attributed to its formation through the knockout of a deeply-bound neutron that breaks the $^{16}$O core and reduces the number of pairs.
\end{abstract}

% \pacs{{\color{red}to be defined}}
\date{\today}
\maketitle

\emph{Introduction} - Pairing correlations play a crucial role in atomic nuclei and quantum many-body physics \cite{Brog13}. In finite nuclei, two-neutron and/or two-proton pairing are responsible for the odd-even staggering observed in the binding energy of atomic masses and for the fact that all even nuclei have a $J^\pi=0^+$ ground state. Pairing correlations also imply a smoothing of the level occupancy around the Fermi energy surface, an enhancement of pair transfer probabilities (see e.g.~\cite{Oer01,Mont14}), as well as a superfluid behavior in nuclear rotation \cite{Migd59} and vibration \cite{Mott75}.
When moving from the interior to the surface of the neutron-rich nuclei $^{11}$Li \cite{Hag07}, $^{6}$He and $^{18}$C \cite{Hag08}, a transition from BCS (Bardeen Cooper-Schrieffer) \cite{BCS} to BEC (Bose-Einstein Condensation) \cite{BEC} pairing has been predicted to possibly occur. 

Tremendous efforts have been made during the last decades to extract information on proton pair correlations from %the decay of
two-proton emitters \cite{Blank08,Pfut12,Blan00,Giov07,Miem07,Asch11} and from the decays of the unbound $^{6}$Be \cite{Gri09,Egor12}, $^{12}$O \cite{Azha98,Jager12}, $^{15}$Ne \cite{Wame14}, $^{16}$Ne \cite{Mukh08,Brown14} and $^{19}$Mg \cite{Mukh08}. While the characterization of the decay (direct or sequential) and structural information on the proton orbitals  involved were obtained with increasing accuracy over the years, all $2p$ decay patterns are subject to strong Coulomb final-state interactions (FSI) that blur the observation of pair correlations at low relative energies.

To circumvent the effects of the Coulomb interaction, the study of two-neutron emission was carried out in neutron-rich core+$n$+$n$ systems that are unbound either in their ground state ($^{10}$He \cite{Joha10}, $^{13}$Li \cite{Joha10,Kohl13}, $^{16}$Be \cite{Spyr12} and $^{26}$O \cite{Lund12,Caes13,Kond16}) or in excited states beyond the two-neutron threshold ($^{8}$He \cite{Laur16}, $^{14}$Be \cite{Marq01,Aksy13} and $^{24}$O \cite{Hoff11, Jone15}).
The decay of excited states of $^{8}$He, $^{14}$Be and $^{24}$O, as well as the ground-state decay of $^{10}$He, all show very convincing signatures of sequential decay through intermediate core-$n$ resonances.
First observations of a di-neutron decay from the ground states of $^{13}$Li \cite{Kohl13} and $^{16}$Be \cite{Spyr12} were claimed on the basis of the observed small $n$-$n$ energies and angles, as compared to a three-body phase-space decay in which the emitted neutrons are free of any interaction.
However, the need to go beyond the di-neutron simplification and use realistic $n$-$n$ FSI, in direct and/or sequential decays, has been pointed out in \cite{Marq12b}. Indeed, the attractive nature of the $n$-$n$ interaction gives rise to small relative $n$-$n$ energies and angles, hereby potentially mimicking a di-neutron decay.

In this Letter, we use the high-energy nucleon knockout reactions $^{19}$N$(-1p)^{18}$C$^*$ and $^{21}$O$(-1n)^{20}$O$^*$ as a `piston' to suddenly promote neutron pairs of $^{18}$C and $^{20}$O respectively into the $^{16}$C+$n$+$n$ and $^{18}$O+$n$+$n$ continuum. Dalitz plots and correlation functions are used to analyze triple correlations in these systems over a decay energy up to 15~MeV above the corresponding two-neutron emission thresholds. An attempt is made to link these observables to the role of the reaction mechanism and to the configurations of $^{18}$C and $^{20}$O,  where the four neutrons above the $^{14}$C and  $^{16}$O cores may be coupled in pairs or tetraneutron configurations \cite{Marq12,Kisa16}.

\emph{Experimental setup} - A stable beam of $^{40}$Ar, accelerated at the GSI facility at 490~MeV/nucleon, was sent on a 4~g/cm$^2$ Be target to induce fragmentation reactions, in which the $^{19}$N and $^{21}$O secondary beams were produced at 430~MeV/nucleon. They were selected by the FRagment Separator \cite{Geis92} and transmitted to the R3B-LAND beam line \cite{Auma07}, in which they were identified using their energy loss and time of flight prior to impinge on a 922~mg/cm$^2$ CH$_2$ reaction target.
The latter was surrounded by the 4$\pi$ Crystal Ball \cite{Metag83}, that detected in-flight photons ($\varepsilon_\gamma\!\sim\!60\%$ around 2~MeV) and protons emitted during the knockout reactions. Two pairs of double-sided silicon strip detectors were placed before and after the target to determine the energy loss and track the incoming and outgoing nuclei.
Nuclei from knockout reactions were deflected by the large dipole magnet ALADIN, and two further position measurements using scintillating fiber detectors allowed for their tracking through the dipole field. The combination with time-of-flight and energy-loss measurements provides the magnetic rigidity and atomic number of the fragments, and therefore their mass and momentum.

% Unbound states in $^{18}$C and $^{20}$O, produced through knockout reactions, emitted neutrons that were detected in the forward direction using the large area neutron detector LAND \cite{Blai92}, positioned 12~m downstream of the reaction target and covering forward angles up to 79~mrad. The $1n$ and $2n$ efficiencies are of the order of 90\% and 70\% up to about 4 and 8~MeV decay energy, respectively, and decrease smoothly beyond those values \cite[Figs.~1,4]{Caes13}.
% The $2n$ efficiency, that includes causality conditions for the rejection of cross-talk events (misidentified $2n$ events induced by a single neutron), drops below 200~keV as neutrons are emitted within a very narrow cone and cannot be distinguished. The energy resolution ($\sigma$) of the observed neutron resonances degrades slowly from 200~keV at 500~keV to 700~keV at 5~MeV excitation energy \cite{Caes13}.
Unbound states in $^{18}$C and $^{20}$O, produced through knockout reactions, emitted neutrons that were detected in the forward direction using the large area neutron detector LAND \cite{Blai92}, positioned 12~m downstream of the reaction target and covering forward angles up to 79~mrad. The energy resolution of the unbound states degrades slowly with increasing decay energy \cite[Fig.~2]{Caes13}. The $1n$ and $2n$ efficiencies are of the order of 90\% and 70\% up to about 4 and 8~MeV decay energy, respectively, and decrease smoothly beyond those values \cite[Figs.~1,4]{Caes13}.
The $2n$ efficiency, that includes causality conditions for the rejection of cross-talk events (misidentified $2n$ events induced by a single neutron), drops below 300~keV as neutrons are emitted within a very narrow cone and cannot be distinguished.

\begin{figure}[t]
\includegraphics[width=0.71\columnwidth]{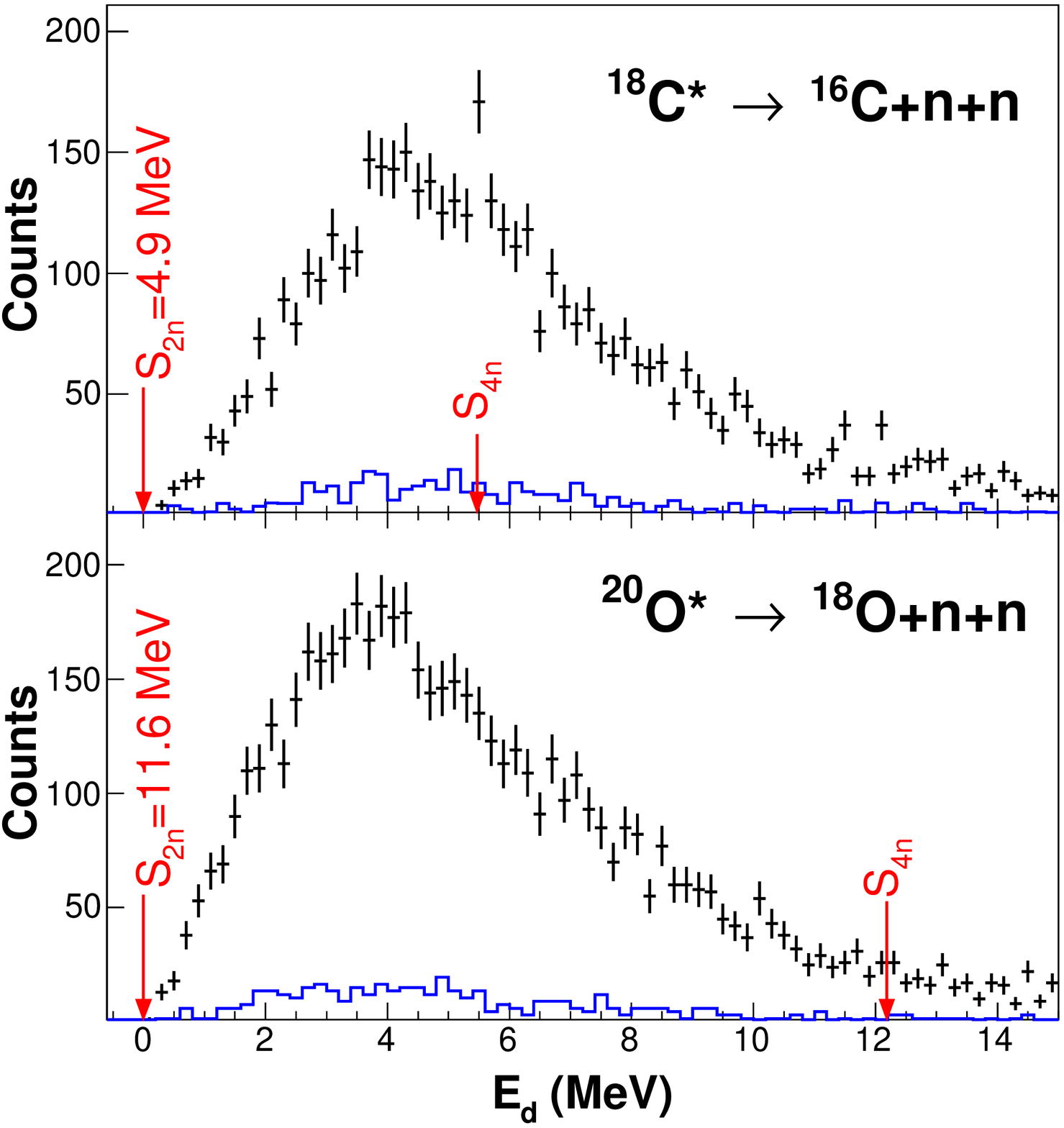}
\begin{minipage}[b]{0.195\columnwidth} \includegraphics[width=\columnwidth]{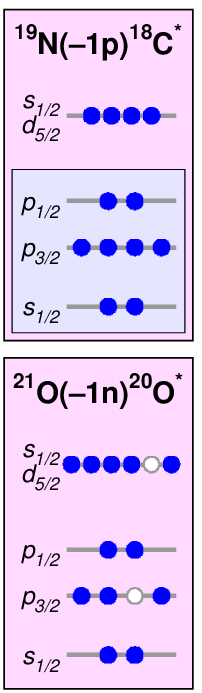} \vspace*{3.4mm} \end{minipage}
\caption{Experimental decay energy spectra of $^{16}$C+$n$+$n$ and $^{18}$O+$n$+$n$ measured respectively in the proton/neutron knockout reactions from $^{19}$N/$^{21}$O 
(blue histograms represent events in coincidence with known $\gamma$-rays in $^{16}$C/$^{18}$O, corrected by $\varepsilon_\gamma$).
The corresponding locations of the $2n$ and $4n$ thresholds are noted. The right panels illustrate the shell model configuration of the 12 neutrons in each isotone.}%10 valence neutrons beyond the $s_{1/2}$ orbit. }
\label{Ed} 
\end{figure}

\emph{Excitation energies} - The invariant mass $M_{fnn}$ of the fragment+$n$+$n$ three-body system, that is reconstructed from the momentum vectors of the fragment and neutrons, is used to calculate  the decay energy $E_d$ of the system ($E_d\!=\!M_{fnn}\!-\!m_f\!-\!2m_n$) in Fig.~\ref{Ed}. 
This energy corresponds to the excitation energy of the total system beyond the $2n$ threshold, since no significant excitation of the fragment (blue histogram in Fig.~\ref{Ed}) has been observed. 
The $2n$-emission spectra of the two nuclei are peaked at about the same energy of 4--5~MeV, and energies up to about 15~MeV were observed. This range of decay energies corresponds to $E^*(^{18}\mbox{C})\!\approx\,$5--20~MeV and $E^*(^{20}\mbox{O})\!\approx\,$12--27~MeV. To reach such high excitation energies, deep nucleon knockout must have occurred. 

At high beam energy, the deep proton knockout reaction $^{19}$N$(-1p)$ is expected to occur mainly through a quasi-free mechanism \cite{Auma13} and \emph{preserve} the structure of the neutrons in $^{18}$C, that can be viewed as a core of $^{14}$C plus 4 neutrons in the $sd$ shells (top-right panel of Fig.~\ref{Ed}). This is supported by the fact that, even if the $^{14}$C threshold is 5.5~MeV higher than the $^{16}$C one (Fig.~\ref{Ed}), the former exhibits a higher yield ($\sigma_{^{14}\rm{C}}/\sigma_{^{16}\rm{C}}\!\sim\!1.8$). This reaction is therefore used here as a tool to suddenly promote neutrons to the continuum, observe their decay, and trace back how they were correlated in $^{18}$C. By contrast, the deep neutron knockout reaction $^{21}$O$(-1n)$ leaves a broken $^{16}$O core and two unpaired neutrons in the $^{20}$O residue (bottom-right panel). %Here, contrary to $^{18}$C, we expect to \emph{favor} configurations in which pairing interactions are reduced, 
In this case, we expect to \emph{hinder} the role of pairing interactions, as will be discussed in view of the present observations.

\begin{figure}[t]
\includegraphics[width=\columnwidth]{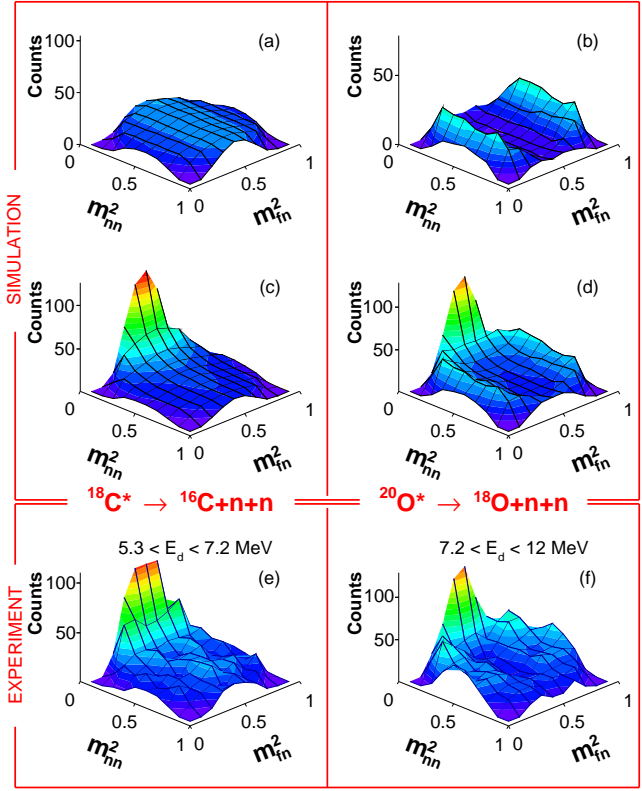}
\caption{Dalitz plots of fragment+$n$+$n$ decays (fragment-$n$ vs $n$-$n$ normalized squared invariant masses). Left panels correspond to $^{16}$C+$n$+$n$, right panels to $^{18}$O+$n$+$n$. The four upper panels represent simulations of (a) phase space, (b) sequential decay through a fragment-$n$ resonance, (c) direct decay with $n$-$n$ FSI, and (d) a combination of the latter two. The lower panels (e,f) correspond to the experimental data for the decay energies noted.} \label{Dalitz} 
\end{figure}

\emph{Dalitz plots} - Correlations in a three-body decay are easily revealed in Dalitz plots of the squared invariant masses of particle pairs ($M^2_{ij}$). FSI and resonances lead to a nonuniform population of those plots within the kinematic boundary defined by energy-momentum conservation and the decay energy \cite{Dalitz}. As our systems are created with a distribution of decay energies, it is convenient to normalize $M^2_{ij}$ between 0 and 1 ($m^2_{ij}$) \cite{Marq01}, so that all events can be displayed within the same boundary, independently of $E_d$. The simulations shown in Figs.~\ref{Dalitz}(a--d) display various correlation patterns as a function of the fragment-$n$ and $n$-$n$ invariant masses, using a model that will be described below. 

In the absence of any correlation beyond phase-space kinematics (a), the plot exhibits a relatively uniform population. If a fragment-$n$ resonance were formed (b), leading to a sequential decay, a band appears at the corresponding value of $m^2_{fn}$, that depends on the resonance energy with respect to $E_d$ (and at $1\!-\!m^2_{fn}$, since $m^2_{fn_2}\!\approx\!1\!-\!m^2_{fn_1}$). The direct decay of a neutron pair induces a concentration of events at $m^2_{nn}\!\lesssim\!0.5$ (c), reflecting the attractive $n$-$n$ interaction.
If the two decay modes coexist (d), a crescent-shaped pattern with a dip at the center appears. Prior to comparing in detail with any model, we can already note that the experimental plot of panel (e) looks almost exclusively like a direct decay, while that of panel (f) displays a mixture of direct and sequential decays. 

\begin{figure}[t]
\includegraphics[width=\columnwidth]{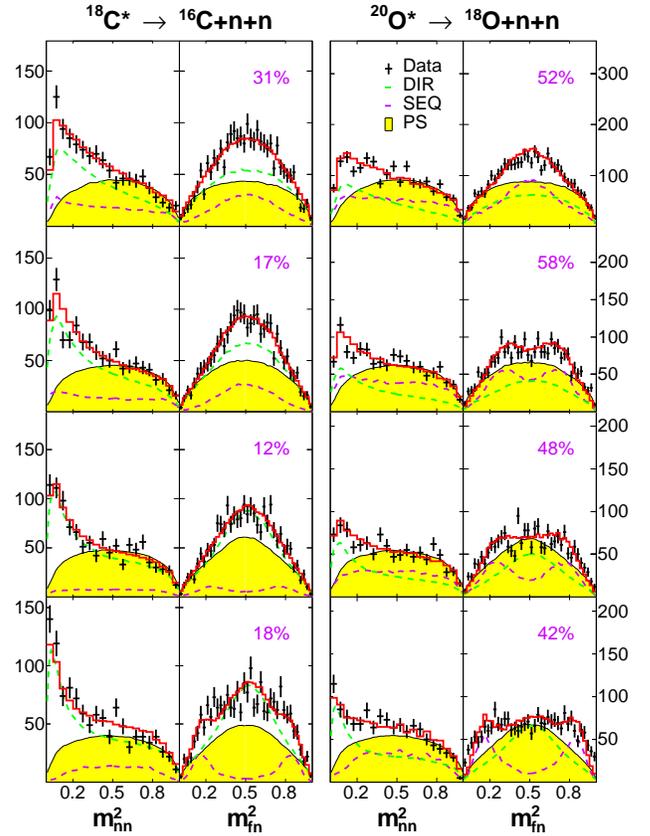}
\caption{Projection of the Dalitz plots defined in Fig.~\protect\ref{Dalitz} onto either axes for the data of $^{18}$C$^*$ (left) and $^{20}$O$^*$ (right) decays. The rows correspond to the four $E_d$ bins defined in the text, from lower (top) to higher (bottom). The yellow histograms represent phase space, normalized to the data at $m^2_{nn}\!>\!0.6$. The red histograms are the projections of the best two-dimensional fit of the plots, with their direct (green) and sequential (purple, with percentage noted) decay components.} \label{fits} 
\end{figure}

The projections of the experimental Dalitz plots are shown in Fig.~\ref{fits} for the two systems and four $E_d$ bins: 0--3.7, 3.7--5.3, 5.3--7.2 and 7.2--12~MeV (chosen in order to contain similar statistics). The phase-space uniform population of the Dalitz plot leads to bell-shaped projections (yellow histograms) with a maximum at about 0.5. They have been normalized to the data at $m^2_{nn}\!>\!0.6$, where no $n$-$n$ correlations are observed. Clearly, the data deviate significantly from phase space. In particular, an increase towards $m^2_{nn}\!=\!0$ is noticeable in all panels, as already observed in Fig.~\ref{Dalitz}(e,f). It is however much stronger in the $2n$ decay of $^{18}$C, which suggests stronger pairing correlations in this system.

Concerning the fragment-$n$ channel, which should reveal the degree of sequentiality in the decay, the expected bands in the Dalitz plot of Fig.~\ref{Dalitz}(b) correspond to `wings' in the projection onto $m^2_{fn}$. Those are clearly observed at 0.1--0.3 and 0.7--0.9 in the three higher-energy bins of $^{20}$O. These wings and the increase of $m^2_{nn}$ towards 0 suggest, 
as was noted above, %in the discussion of Fig.~\ref{Dalitz}(f),
 that the sequential and direct decays are in competition.
% Determining to what extent they compete will be done in comparison to a model that contains both components.
In order to determine the extent of this competition, we have used a phenomenological model that contains both components.

\emph{Correlation functions} - The interaction effects within a pair of particles are, by definition, best displayed through the correlation function $C$. It represents the ratio of the measured two-particle distribution and the product of the independent single-particle ones, that those particles would exhibit without their mutual influence \cite{Marq00}. For most particle pairs the correlation signal, including the effects of FSI and, for identical particles, quantum statistics, manifests at low relative momenta $q_{ij}\!=\!|\vec{p}_{i}\!-\!\vec{p}_{j}|$ \cite{Led82}. In the case of bosons, charged fermions or long time scales, the signal at zero relative momentum is weak, $C(0)\!\ll\!2$ \cite{Boal90,Marq97,Colo95}. For neutrons, however, the attractive FSI may lead to values as high as $C(0)\!\sim\!10$--15 \cite{Marq00}.

\begin{figure}[t]
\includegraphics[height=5.5cm]{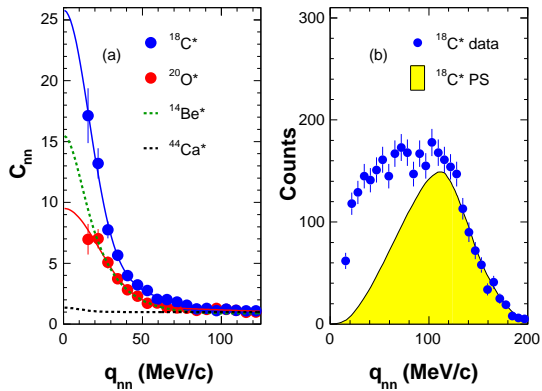}
\caption{(a) Two-neutron correlation functions from the three higher-energy bins of $^{18}$C$^*$ (blue) and $^{20}$O$^*$ (red) $2n$ decays. The solid lines are traced to guide the eye, while the dashed lines correspond to the fits of the experimental data from the breakup of $^{14}$Be (green) \protect\cite{Marq01} and the neutron evaporation from $^{44}$Ca (black) \protect\cite{Colo95}. (b) Numerator (measured relative momentum  distribution, blue points) and denominator (phase space, yellow) of $C_{nn}$ for the $^{18}$C$^*$ case.}%, with the fit from panel (a) multiplied by phase space (blue solid line).}
\label{Cnn} 
\end{figure}

The experimental correlation functions $C_{nn}$ of Fig.~\ref{Cnn}(a) have been constructed for $^{18}$C (blue dots) and $^{20}$O (red dots) from the ratio of the measured relative momentum  distribution $q_{nn}$, that contains the interaction effects, and the one obtained from phase space, that contains all other effects like kinematic constraints or the experimental filter. These two distributions are shown in Fig.~\ref{Cnn}(b) for the $^{18}$C case, where the effect of the $n$-$n$ FSI at  $q_{nn}$ values below 100 MeV/$c$ becomes even clearer.
In order to guide the eye, the experimental $C_{nn}$ have been fitted with two Gaussians. The correlation signal in $^{18}$C, $C_{nn}(0)\!\sim\!25$, is huge, actually the largest ever observed.

In order to interpret this correlation strength, the authors of Ref.~\cite{Led82} propose a formulation that links $C_{nn}(q_{nn})$ to the size and lifetime of a Gaussian source emitting independent neutrons. When the source of particle pairs is large and/or the emission of the two particles proceeds through a long decay time, correlations are expected to be very weak. Within this formalism, the $^{18}$C data would suggest a small source and a very short decay time, or a very weak contribution of the sequential decay, as was anticipated already in Fig.~\ref{Dalitz}(e).

For comparison, we have added in Fig.~\ref{Cnn}(a) the correlation functions obtained for two significantly different systems. In one case (black dashed line), the source of neutron pairs was the compound nucleus formed in the collision $^{18}$O+$^{26}$Mg \cite{Colo95}. The best fit of the experimental $C_{nn}$ was obtained for a sphere of $R\!=\!4.4\pm0.3$~fm %(corresponding to $\rms\!=\!5.3\pm0.4$~fm) 
and a lifetime of $\tau\!=\!1100\pm100$~fm/$c$. For this moderately small source, the long decay time scale is responsible for shrinking the correlation to $C_{nn}(0)\!\sim\!1.3$, a signal about a factor 80 smaller than the one measured for $^{18}$C. 

In the second case (green dashed line), the source was formed during the breakup of the two-neutron halo nucleus $^{14}$Be \cite{Marq01}. Direct pair emission ($\tau\!=\!0$) was invoked to account for the strong correlation measured, $C_{nn}(0)\!\sim\!15$, at that time the largest ever observed. However, the relatively large size of the neutron pair in this halo nucleus, with a correlation signal described by a Gaussian source of $\rms\!=\!5.6\pm1.0$~fm, accounts for a reduction of about 40\% with respect to $^{18}$C.

%As a first step, the solid lines in Fig.~\ref{Cnn} are a fit with the formula \cite[Eq.~(24)]{Led82}, that becomes analytical for $\tau\!=\!0$. This parametrization, that would lead to an effective distance, is used here just to guide the eye. We find a good description of the experimental $C_{nn}$ for both $^{18}$C and $^{20}$O, even if within this formalism internal momentum correlations in the source are neglected. In the following, we will use the explicit ($\rms,\tau$) parametrization of Ref.~\cite{Led82} and assume that, although our neutrons are not independent, those potential correlations in the source are small or have a negligible impact on $C_{nn}$ after averaging over the whole distribution. This assumption, together with the suitability of using a Gaussian source, has been discussed in detail in Ref.~\cite{Laur16}.

\emph{Decay model and results} - In order to include the different correlations observed above phase space, we have used the model developed in Ref.~\cite{Marq01}.
This model does not include the microscopic structure of the initial state, and treats the effects of FSI and resonances on the fragment+$2n$ phase-space decay phenomenologically (for a detailed discussion of its applicability, see Ref.~\cite{Laur16}).
In brief, the experimental decay energy distributions of Fig.~\ref{Ed} are used to generate events with $\vec{p}_f,\vec{p}_{n_1},\vec{p}_{n_2}$ following either three-body phase space (direct decay), or twice the two-body phase space through a fragment-$n$ resonance (sequential). In the latter case, a neutron and the fragment-$n$ resonance are generated first, followed by the decay of the resonance. In both cases, the $n$-$n$ FSI is introduced via a probability $P(q_{nn})$ with the form of the $n$-$n$ correlation function \cite{Led82}, that depends on the space-time parameters ($\rms,\tau$) of a Gaussian two-neutron source.

In an attempt to reduce the parameters of the fit to a reasonable number, we consider that the sequential decay occurs through one fragment-$n$ resonance of energy $\langle{E_R}\rangle$ and width $\langle{\Gamma_R}\rangle$, that can be seen as an average over individual resonances. In fact, even the fits of the higher-energy bins only require one low-energy resonance, of $\langle{E_R}\rangle\!\sim\!1.5$~MeV, like in Fig.~\ref{Dalitz}(b,d). 
The number of free parameters, $\rms$, $\tau$, fraction of sequential decays, $\langle{E_R}\rangle$ and $\langle{\Gamma_R}\rangle$, are further reduced by equating the delay induced in the neutron emission with the lifetime of the fragment-$n$ resonance, leading to $\tau\!=\!\hbar{c}/\langle{\Gamma_R}\rangle$. This was demonstrated in Ref.~\cite{Laur16} for the well-known $^7$He resonance, although in the present case the average over several resonance energies might lead to an effective delay that does not correspond well with the individual lifetimes.

The final momenta of the three generated particles are filtered to include all experimental effects (like energy resolution, angular acceptance, or cross-talk rejection). Then the different observables are reconstructed and subsequently fitted to the data in the two-dimensional Dalitz surface (Fig.~\ref{Dalitz}), with a combination of direct and sequential decay modes. An example of the goodness of the two-dimensional fit is given in the comparison between panels (d) and (f), where both the $n$-$n$ FSI and the wings of the sequential mode are accurately reproduced. Similar agreement is found for all the Dalitz plots (not shown here) as well as for their projections shown in Fig.~\ref{fits}, further validating the different hypotheses used. 

% {\color{red} From the fits at the higher-energy bin (in which the fraction of sequential decay is the best constrained), we derive a compact configuration of $3.7\pm0.15$~fm in both systems.} An average $\rms$ over the three higher-energy bins leads to $4.1\pm0.7$~fm for $^{18}$C and $4.2\pm0.9$~fm for $^{20}$O, in line with the value corresponding to independent neutrons in a sphere of radius $1.2\,A^{1/3}$ (4~fm for $A\!=\!20$). The stronger $n$-$n$ signal in $^{18}$C is due to the fact that the neutron pair is emitted directly in (84$\pm$12)\% of the time. The sequential branch only becomes slightly apparent in the wings of the highest-energy bin (Fig.~\ref{fits}). In contrast, (49$\pm$14)\% of the decays are sequential in $^{20}$O, with wings in $m^2_{fn}$ that are visible in all bins, even in the lowest energy one in which they move towards 0.5 to create an enhanced central contribution there. 

Considering the average over the four energy bins, the fits denote a compact configuration in both systems, corresponding to a Gaussian source of $\rms\!=\!4.1\pm0.4$~fm for $^{18}$C and $4.3\pm0.6$~fm for $^{20}$O. Both values are in line with the one corresponding to independent neutrons in a liquid drop of $A\!=\!20$ (4~fm). According to the fits, however, the stronger $n$-$n$ signal in $^{18}$C is due to the neutron pair being emitted directly in 81$\pm$9\% of the time, with a sequential branch only slightly apparent in the wings of the highest-energy bin. In contrast, 50$\pm$8\% of the decays are sequential in $^{20}$O, with wings in $m^2_{fn}$ that are visible in all bins, even in the lowest energy one in which they move towards $m^2_{fn}\!=\!0.5$ to create an enhanced central contribution there.
% From these fits, we derive a compact configuration in both systems, with an average $\rms$ over the three higher-energy bins of $4.1\pm0.7$~fm for $^{18}$C and $4.2\pm0.9$~fm for $^{20}$O, in line with the value corresponding to independent neutrons in a sphere of radius $1.2\,A^{1/3}$ (4~fm for $A\!=\!20$). The stronger $n$-$n$ signal in $^{18}$C is due to the neutron pair being emitted directly in (84$\pm$12)\% of the time. The sequential branch only becomes slightly apparent in the wings of the highest-energy bin (Fig.~\ref{fits}). In contrast, (49$\pm$14)\% of the decays are sequential in $^{20}$O, with wings in $m^2_{fn}$ that are visible in all bins, even in the lowest energy one in which they move towards $m^2_{fn}\!=\!0.5$ to create an enhanced central contribution there. 

\emph{Conclusions} - High-energy nucleon knockout reactions have been used to populate unbound states in the $N\!=\!12$ $^{18}$C and $^{20}$O isotones up to 15~MeV above their two-neutron emission thresholds.
Their three-body decay was characterized by the combined determination of the momenta of the residual fragment %, detected in the ALADIN spectrometer,
and the two neutrons. The experimental fragment-$n$ and $n$-$n$ invariant masses have been compared to those obtained from a three-body decay model that takes into account direct and sequential decays, as well as final-state interactions.%

The decay of the core+$4n$ isotones $^{18}$C and $^{21}$O displays significantly different features. In the former, extremely strong correlations persist up to 12~MeV, which we propose to be caused by the large fraction ($\sim$\,80\%) of direct emission of correlated pairs with a relatively compact configuration. The decay of $^{20}$O exhibits much weaker correlations, with about 50\% occurring through sequential processes. The clear contrast between these isotones is likely due to the way they are populated: the knockout of deeply-bound neutrons from $^{21}$O leaves two unpaired neutrons in $^{20}$O with a broken $^{16}$O core (in this way increasing the probability of sequential decay), while the knockout of deeply-bound protons from $^{19}$N leaves the neutron pairs and the $^{14}$C core unaffected. 

The present study shows that the high-energy proton knockout reaction is a tool of choice for studying neutron correlations, be there of $2n$ or $4n$ origin, up to the neutron drip line. It is hoped that the present results will encourage theoretical calculations to interpret the present experimental observables on a more microscopic ground, similar to those employed in proton-rich systems \cite{Gri09,Egor12}.

\acknowledgments {\small
A. Chbihi, S. Gal\`es,  J.-P.~Ebran and  L.~Sobotka, are greatly acknowledged for fruitful discussions. 
This work was supported by the German Federal Ministry for Education and Research (BMBF project 05P15RDFN1), and through the GSI-TU Darmstadt co-operation agreement.
C.A.~Bertulani acknowledges support by the U.S.\ DOE grants DE-FG02-08ER41533 and the U.S.\ NSF Grant No.~1415656.}

%\newpage

%\end{multicols}

\end{document}